\documentclass{PoS}
\usepackage{graphicx}

\def\hi {H\,{\sc i~}}

\def\kms{km\,s$^{-1}$}
\def\kmss{km\,s$^{-1}~$}

\def\deg{\hbox{$^\circ$}}

\title{A panoramic view of the Milky Way HI gas}

\ShortTitle{A panoramic view of the Milky Way HI gas}

\author{\speaker{Peter M.W. Kalberla},$^a$ Naomi M. McClure-Griffiths$^b$
  \thanks{For the GASS consortium.}, J\"urgen Kerp$^a$ \thanks{For the EBHIS consortium.} \\
  \llap{$^a$}Argelander-Institut f\"ur Astronomie, Auf dem H\"ugel 71, D-53121
  Bonn, Germany\\
  \llap{$^b$}Australia Telescope National Facility, CSIRO, Marsfield
  NSW 2122, Australia\\

  E-mail: \email{pkalberla@astro.uni-bonn.de},
    \email{Naomi.Mcclure-Griffiths@csiro.au}, \email{jkerp@astro.uni-bonn.de}}

\abstract{Neutral atomic hydrogen (\hi) traces the interstellar medium (ISM)
  over a broad range of physical conditions. Its 21-cm emission line is a key
  probe of the structure and dynamics of galaxies. This line comprises very
  different temperature and density regimes on all scales, from tens of
  astronomical units to kiloparsecs. \hi is the key element to study the
  evolution of the ISM in detail. To understand the physics of the ISM and to
  analyze the interplay between different phases it is mandatory to cover
  observationally a broad range of scales. But large scale imaging of
  galaxies, resolving at the same time all these scales is difficult; spatial
  resolution, as well as sensitivity and the field of view are currently
  rather limited.

  The observational situation is much more favorable if we consider our own
  galaxy. Two major all sky 21-cm line surveys of the Milky Way will become
  available soon. The Galactic All Sky Survey (GASS) obtained with the Parkes
  64-m telescope\thanks{The Parkes Radio Telescope is part of the Australia
    Telescope which is funded by the Commonwealth of Australia for operation
    as a National Facility managed by CSIRO}~ for the southern hemisphere with
  a resolution of 16 arcmin is close to completion. The northern extension,
  the Effelsberg Bonn HI Survey (EBHIS)\thanks{Based on observations with the
    100-m telescope of the MPIfR (Max-Planck-Institut f\"ur Radioastronomie)
    at Effelsberg}~ with 9 arcmin resolution, will be available in 2010/2011;
  we refer to the talks by Kerp and Winkel. Here we discuss briefly the GASS
  and demonstrate the unprecedented quality of this survey.

  The Galactic single dish 21-cm line surveys prepare the ground for future
  high resolution imaging of the Galactic \hi distribution. Using the
  available short spacing informaton, the Australian Square Kilometre Array
  Pathfinder (ASKAP) will be capable to generate a
  truly panoramic view of the Milky Way HI gas distribution with arcsecond
  resolution for all declinations $ < 30\deg$. Data from the Widefield ASKAP
  L-band Legacy All-Sky Blind Survey (WALLABY, see talk by Staveley-Smith) can
  be used to generate high resolution all sky maps. In comparison to the
  currently available interferometric International Galactic Plane Surveys
  (IGPS) the sensitivity will improve by a factor of 10. Most important is the
  all sky coverage which will overcome the rather limited spatial coverage of
  a few degrees around the Galactic plane for the IGPS.}

\FullConference{Panoramic Radio Astronomy: Wide-field 1-2 GHz research on galaxy evolution\\
		 June 2-5 2009\\
		 Groningen, the Netherlands}

\begin{document}

\section{The Milky Way galaxy, an outstanding source for ISM research}
Several major Galactic 21-cm line surveys have been released during the past
decade \cite{ARAA47}. Based on this our knowledge about details in the \hi
distribution in our Galaxy is far better than that of other galaxies. To
demonstrate observational limitations in galactic research we use M31 as an
example. Observing this galaxy with a 1 arcmin beam, we can resolve a linear
dimension of 200 pc. The situation for more distant galaxies is even
worse. Such scales are barely of interest for the physics of the interstellar
medium \cite{ARAA47}.

For the Milky Way the situation is much more favorable. For a 1 arcmin beam
the resolution in comparison to M31 is better by at least a factor of 20 for
even the most distant regions of the Milky Way. The spatial resolution
obtained with large single dish telescopes is comparable to interferometric
observations of M31 but in this case the single dish sensitivity is better by a
factor of 20. Large scale surveys reach easily a sensitivity of 50
mK which is quite a challenge for synthesis telescopes. 

This simplistic comparison demonstrates that the Milky Way galaxy, from
observations, is an
outstanding source for ISM research. In the following we give first results
for the GASS and argue for supplementing single dish observations with high
resolution ASKAP observations.

\section{GASS: a high resolution, sensitive, and accurate 21-cm line survey}

The Galactic All Sky Survey was observed with the Parkes 64-m telescope and
covers declinations $ \delta < 1\deg$. The GASS is the most sensitive, highest
angular resolution survey of Galactic \hi in the Southern sky
\cite{GASS1}.  

Single dish observations of the Galactic HI line emission
suffer from stray radiation originating at the rim of the primary mirror
(spill-over) and also from reflections at the feed support legs
(stray-cones). It is therefore mandatory to calculate and eliminate this
contribution from the observations. Such corrected GASS data will be available
soon, here we demonstrate the superior quality of the forthcoming second GASS
data release.

Fig. 1 shows the emission at $v = -30$ \kms. The map in the center results
after correcting the observations for stray radiation and other instrumental
effects. The contribution from the beam pattern which was calculated by
convolving the antenna sidelobes with the \hi emission on the sky and which
was subtracted from the observations is shown on top of Fig. 1. The obvious
patchy structure results from sidelobe contributions that vary with time and
season, the patches correspond to individual observing sessions. The line
emission after cleaning (center panel) shows no correlations with the
observing patches and discloses a rich wealth of faint filaments which were
previously hidden. The bottom panel displays data from the
Leiden/Arentine/Bonn (LAB) survey \cite{LAB}. These data have also been
corrected for instrumental biases. Previously the LAB was considered to be the
most accurate 21-cm line survey but apparently not all of the stray radiation
could be eliminated; boxy artifacts correlate also in case of the LAB with 
observing sessions.

We conclude that the quality of the GASS with a sensitivity of 40 mK is much
superior to all other currently available Galactic 21-cm line surveys of the
Southern sky.

\begin{figure}
\includegraphics[width=.48\textwidth,angle=-90]{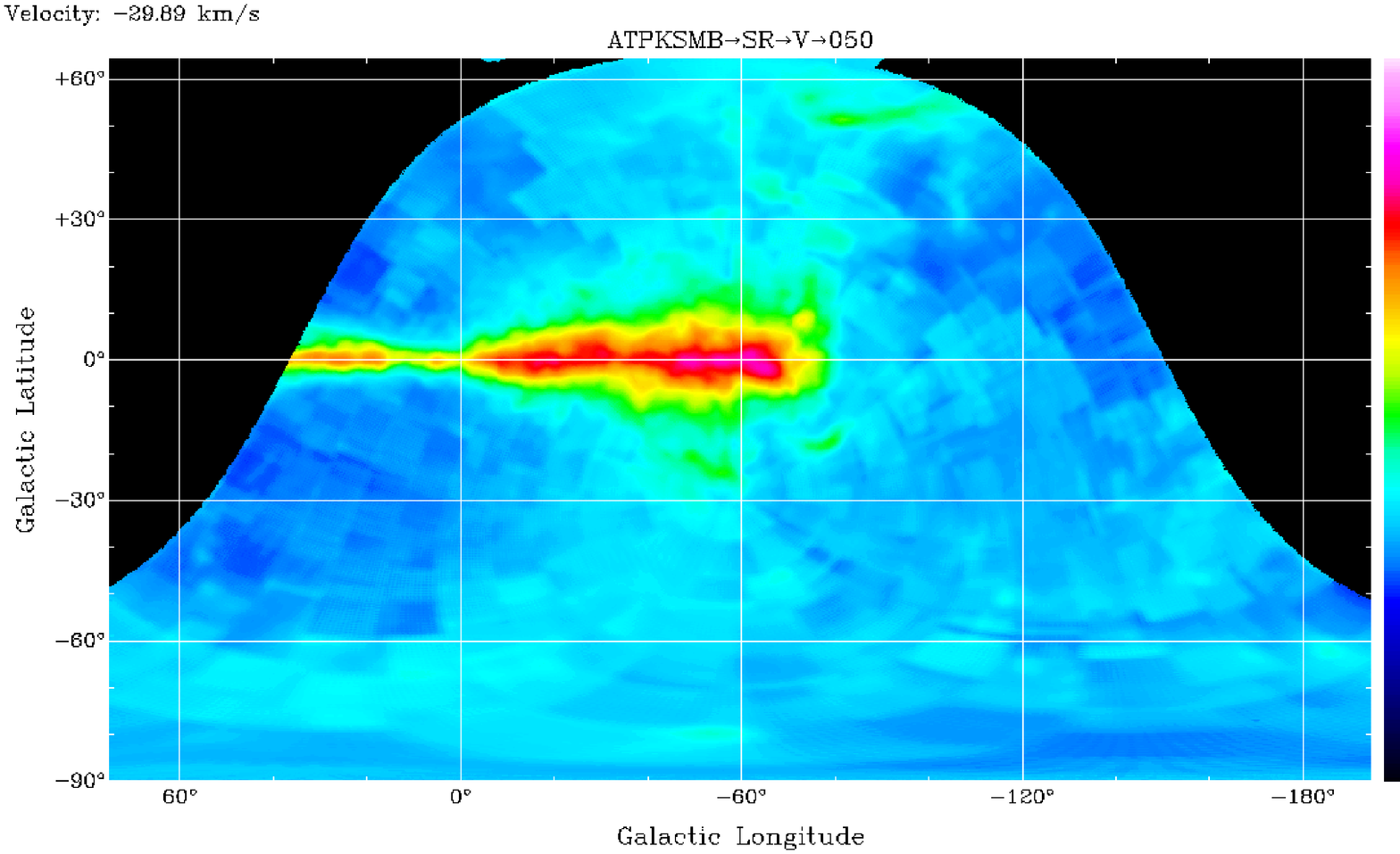}
\includegraphics[width=.48\textwidth,angle=-90]{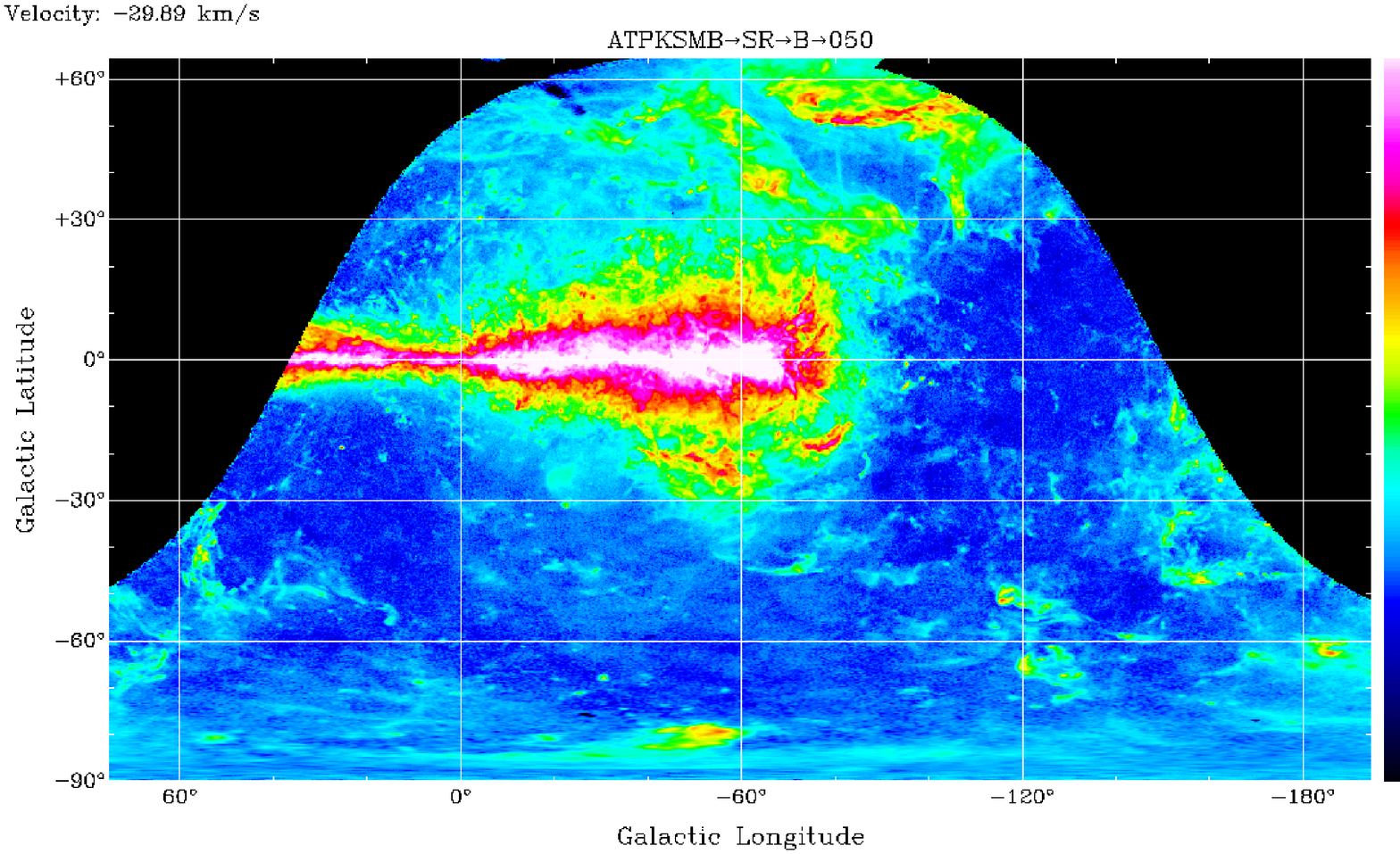}
\includegraphics[width=.48\textwidth,angle=-90]{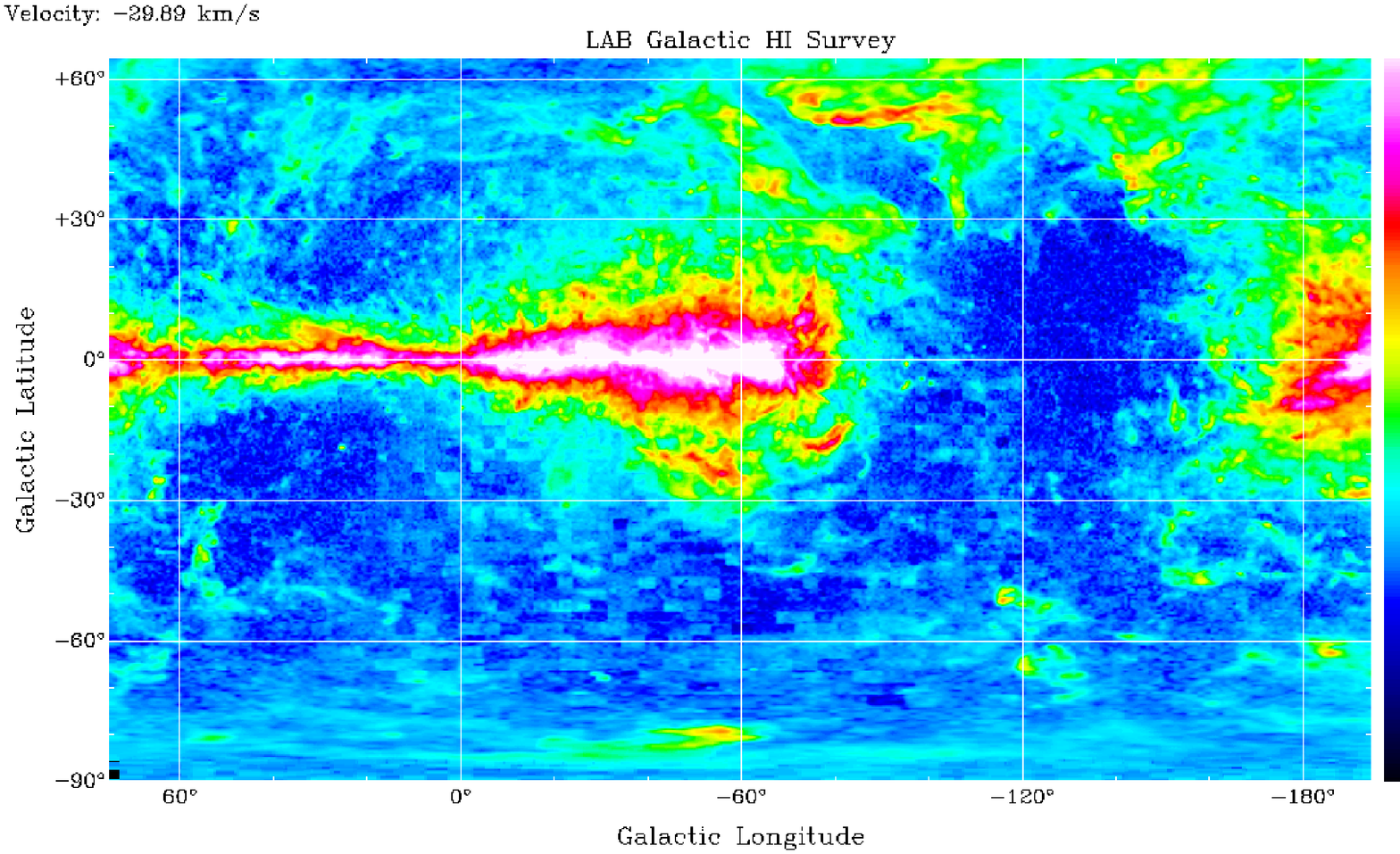}
\caption{Three channel-maps at $v = -30$ \kms. The GASS map is in the middle,
  stray radiation which was eliminated is on the top. At the bottom the LAB
  map is shown for comparison. A logarithmic transfer function was chosen to
  emphasize low brightness temperatures. Black areas indicate positive declinations.}
\label{fig1}
\end{figure}

\section{ASKAP: aiming for panoramic images of the Galactic \hi gas}

Galactic surveys are essential for ISM research and sensitive high resolution
single dish surveys like GASS and EBHIS are a major step forward.  Still,
there is a need for supplementing high resolution data.  The \hi line
comprises very different temperature and density regimes on all scales, from
tens of astronomical units to kiloparsecs. In particular, the analysis of the
cold and clumpy component of the multi phase medium on scales of parsecs and
below requires interferometer observations. This concerns pressure balance
between the different phases and transitions between them. Tiny scale atomic
structures do exist with low column densities and linear dimensions of a few
hundred or thousand astronomical units.  Also little is known about the
disk-halo interface and the regulation of the gas flow. If we want to
understand galaxy formation and evolution we need first to understand the
processes on scales where critical transitions happen.

The new \hi surveys will open a new window to the early universe. Future X-ray
observatories will have most of their detection power in the soft X-ray energy
range below 1\,keV. This is caused by the large photoelectric absorption cross
section of the ISM. The softer the X-ray radiation of interest, the larger the
attenuation of the X-rays by the Galactic interstellar medium. The emission of
active galactic nuclei at high redshifts ($z \sim 10$) or the faint emission
of clusters of galaxies at moderate redshifts ($3 < z < 5$) is shifted to this
soft X-ray band. It is not possible to analyze the X-ray data quantitatively
without knowing in detail the distribution of the Galactic interstellar
medium. High resolution data on scales of arcminutes and below are needed. Our
current view through ``windows'' like the Lockman hole and the Chandra
deep-field south may result in a tunnel vision of the early universe,
strongly biasing our knowledge towards specific large scale structures. 

ASKAP offers a unique chance to supplement the GASS and EBHIS surveys with
high resolution interferometric data. We propose to use data from the
Widefield ASKAP L-band Legacy All-sky Blind surveY (WALLABY, presented by
Staveley-Smith) to obtain a complete unbiased high quality database at
arcsecond resolution.  For a synthesized beam of 60 arcseconds a sensitivity
of 300 mK is feasible, a factor of 10 improvement with respect to the
currently available IGPS data.  Such a Galactic ASKAP survey would cover the
southern sky up to declinations $ \delta < 30\deg$. At arcsecond resolution
the ASKAP images would provide a truly panoramic view of the Galactic \hi gas
over a broad spectral window of $ -400 < v < 400 $ \kmss with a velocity
resolution of 4 \kms.
\\

{\bf ACKNOWLEDGEMENTS:} PK acknowledges support from Deutsche
Forschungsgemeinschaft, grant KA1265/5, JK from grant KE757/7.

\end{document}